\documentclass[a4paper,fleqn]{cas-dc}

\usepackage[numbers]{natbib}
\usepackage{comment}
\usepackage{longtable}

\usepackage[utf8]{inputenc}
\usepackage{url}
\usepackage{xspace}
\usepackage{graphicx}
\usepackage{listings}
\usepackage{setspace}
\usepackage{tabularx}
\usepackage{multirow}
\usepackage{comment}
\usepackage{dirtytalk}
\usepackage{enumitem}

\lstdefinestyle{mystyle}{
  numberstyle=\tiny,
  basicstyle=\footnotesize\ttfamily,
  breakatwhitespace=false,         
  breaklines=true,                 
  captionpos=b,                    
  keepspaces=true,          
  numbers=none,
  numbersep=5pt,                  
  showspaces=false,                
  showstringspaces=false,
  showtabs=false,                  
  tabsize=2,
  frame=tb
}

\def\tsc#1{\csdef{#1}{\textsc{\lowercase{#1}}\xspace}}
\tsc{WGM}
\tsc{QE}
\tsc{EP}
\tsc{PMS}
\tsc{BEC}
\tsc{DE}

\begin{document}
\let\WriteBookmarks\relax
\def\floatpagepagefraction{1}
\def\textpagefraction{.001}
\shorttitle{A holistic approach for cross-platform software development}
\shortauthors{JZ Blanco and D Lucrédio}

\title [mode = title]{A holistic approach for cross-platform software development}                      



\author[1,2]{J. Z. Blanco}[orcid=0000-0002-2530-6917]
\cormark[1]
\ead{julianoblanco@ifsp.edu.br}

\credit{Conceptualization of this study, Methodology, Software, Data curation, Funding acquisition, Project administration, Resources, Writing - Original draft preparation, Validation, Investigation}

\address[1]{J. Z. Blanco and D. Lucrédio are with the Computing Department, Federal University of São Carlos, Rod. Washington Luís, Km~235, P.O. Box 676 - 13565-905, São Carlos - SP - Brazil.}

\author[1]{D. Lucrédio}[orcid=0000-0002-1360-4036
]

\ead{daniel.lucredio@ufscar.br}

\credit{Conceptualization of this study, Methodology, Software, Data curation, Funding acquisition, Project administration, Resources, Supervision, Writing - Original draft preparation}

\address[2]{J. Z. Blanco is with Federal Institute of São Paulo, Campus Piracicaba, Rua Diácono Jair de Oliveira, 1005, 13414-155, Piracicaba - SP - Brazil.}



\cortext[cor1]{Corresponding author}


\begin{abstract}
Cross-platform development solutions can help to make software available on different devices and platforms. But these are normally restricted to preconfigured platforms and consider that each individual solution is equal or similar to each other. As a result, developers have to resort to native development and build individual solutions, one for each device/platform, that cooperate to deliver the desired global functionality. This article presents an approach that takes advantage of existing solutions and have support for extending and including new platforms, and distributing functionality across devices. The approach is based on a general-purpose language that raises the abstraction level in order to keep the software free from platform details. Automatic transformations produce executable code that can be properly divided and deployed separately into different platforms. The proposed approach was evaluated in four ways. In the first evaluation, an existing cross-platform system was recreated using the approach. The second and third evaluations was conducted with expert and novice developers, who tested the approach in practice. The fourth evaluation introduced support for cross-platform testing. Results have brought evidence supporting the following main contributions: use of a single environment, the ability to reuse similar concepts between platforms and the potential to reduce costs. 
\end{abstract}




\maketitle

\section{Introduction}
\label{sec:intro}

The great number and variety of computing devices, such as smartphones, tablets and smartwatches, creates an ever-growing demand for cross-platform software, that is, a single software system that runs on multiple devices. In this scenario, developers are challenged to handle device-specific characteristics such as types and versions of operating systems, data storage capacity, available resources (GPS, camera, etc.), screen size, and more \cite{Wasserman2010}. Dealing with these differences at a low abstraction level have a impact on productivity, maintenance and quality, ultimately resulting in cost increases and loss of focus on the business domain.


There are two additional complications to this scenario. First, there is always the risk of having to support new platforms in the near future. The industry has been investing in the use of computing in various objects \cite{hoffman2018consumer} and this reality should be considered for software engineering. To keep software competitive, in many cases it is important to run on as many new platforms as possible, as soon as possible.

Second, in some cases, the software must have its functionalities distributed between versions for different platforms \cite{7972737}. Not always the functions of one version for a specific platform should be maintained in other versions. For example, a mobile smartphone is more suitable for taking geotagged photographs, while a computer is preferred for typing or viewing large pictures or maps. Functions distributed across system versions can better explore the capabilities of each device.

The problem is not recent. In the recent past, when smartphones started to appear, Java emerged as a promising solution, in the form of the Java 2 Micro Edition. In theory, a software could be written in a single way (using J2ME) and run everywhere, as long as there was a supporting virtual machine. But even in the realm of mobile phones, which is a very well defined context, different implementations were provided by different vendors, causing software to crash unless extensively tested on dozens of devices \cite{umuhoza2015automatic}. Also, each manufacturer started to add specific features, such as dedicated sensors or specific controls. Supporting all of these variations in a single framework was very problematic, and developers had to deal with these differences in the code, by adding preprocessing directives or maintaining completely different versions of the software at the same time \cite{chen2019automated}. Eventually, when devices became powerful enough, J2ME and the promise of Java as a universal mobile language approach was replaced by Android, a complete operating system \cite{Heitkotter2013}.

Current web technologies such as HTML5 
and Progressive Web Apps\footnote{developers.google.com/web/updates/2015/12/getting-started-pwa} can help to make software available in any device with a web browser. It is possible to develop responsive interfaces, suitable for different screen sizes, with offline capabilities and background processing. These can make the experience become closer to a native application, but there are limitations. Accessing hardware features such as Bluetooth, Near-Field Communication, GPS and Camera is not ideal \cite{litayem2015review}. There is also a limiting factor regarding local storage in the browser\footnote{developers.google.com/web/fundamentals/instant-and-offline/web-storage/offline-for-pwa}. Finally, there are devices such as smartwatches, that have very limited hardware and do not feature a fully functioning browser. In these cases, native development is the only option.

There are solutions that help to reduce some of these challenges. Hybrid development \cite{Heitkotter2013} and generative approaches \cite{czarnecki2002generative} can help to unify the different software versions, but they normally have poor support for a variety of current and new platforms. Responsive web development \cite{Marcotte} have good potential to support different screen types but have limitations regarding native hardware \cite{litayem2015review}. Service-oriented architecture and cloud computing \cite{moreno2017cross} may help by moving parts of the system out of the devices and facilitate maintenance of each individual service, but device and platform-specific functions still have to be developed to consume these services. 
None of these solutions solve the essence of the problem, which is the separate development for what should be treated more like a single system. The developers still have to work with multiple individual systems that work together and deal with the technical details inherent to each platform. Also, when adopting a particular solution, the developer is normally restricted to whichever platforms are currently supported by the tool or approach.

In face of these challenges, this article presents an approach where development and maintenance consider a cross-platform system as a single software entity. Therefore, the approach is considered holistic as it covers multiple characteristics and seeks to uniformly solve the challenges associated with cross-platform development. The developer creates the software using a single set of models, using a Generic Purpose Language (GPL) developed to support the approach. The GPL is a programming language with some high level constructs that allow different domain concepts and functions to be specified completely and independently from the platform(s) on which they will run. Through generators, code is automatically obtained for different platforms.

The approach is also holistic in the sense that it is not restricted to a predefined set of supported platforms. Adding new platforms is embedded in the approach through the same GPL, with which platform models can be specified and code be generated. Therefore, the approach can cover a wide range of devices, both current and future ones. Furthermore, the approach allows existing platforms to be extended or modified, to better suit the needs of a particular system.

Finally, the approach can be used to choose in which platform each part or function will be deployed. This facilitates the job of distributing functionality across devices, for example deploying GPS and camera-dependent functions into smartphones only, and larger reports and data sheets functionality into a web application. This is a major difference from many existing solutions, which normally create different versions of the same software for different platforms.

To evaluate the approach, four studies were conducted. A first study was conducted by the researchers, and consisted in a proof-of-concept where an existing commercial cross-platform system was recreated in the platform. A second study involved five experts, who used the approach to create software using the approach. To demonstrate the holistic nature of the approach in supporting different and new platforms, the studies involved deploying the software in the web, Android and iOS, with different storage technologies (HashMap and SQLite), different programming languages (Java, C\# and Swift), and even included a new device created solely for this research: an augmented reality (AR) device based on RaspBerry Pi\footnote{https://www.raspberrypi.org/}, with simple input/output capabilities. The third study was similar to the second, but involved four developers with low expertise in mobile development. And the fourth study explored cross-platform unit testing.


The results show that the approach can be successfully used to treat cross-platform development as a single software entity. It also highlighted some contributions, the main being the possibility to use a single environment to create a platform-independent representation of the software. The results also demonstrated the ability to reuse similar concepts between platforms. Ultimately, the approach has the potential to reduce costs. The evaluations also pointed out some limitations, the most important being the need for an initial effort to properly create and prepare the platform details. This is important, as most developers have this support already available in current tools. Also, in some cases it was necessary to use the platforms' native IDEs, thus breaking the purpose of treating the software as a single entity in a single environment. Another limitation is the need for additional tests, to make sure the generated code is adequate and can be trusted. The results also point out some implementation details that could be improved in the future. The collected empirical evidence constitutes important contributions for continuing this research in the future until it is ready to reach production-ready tools and environments.

The remainder of this paper is organized as follows. Section \ref{sec:relatedWork} presents related work. Section \ref{sec:approach} describes the cross-platform development approach in details. Section \ref{sec:Evaluation} contains the evaluation and Section \ref{sec:conclusion} presents some concluding remarks and future work.

\section{Related Work}
\label{sec:relatedWork}





HAXE\footnote{haxe.org} and KOTLIN\footnote{kotlinlang.org} are two examples of languages that can be used to create platform-independent software by generating code for different native languages, including JavaScript, PHP, Pyhton, C++, Java, among others, offering support for different platforms, including Windows, Linux, MacOS, iOS and Android.
The IBM RATIONAL RHAPSODY\footnote{www-03.ibm.com/software/products/en/ratirhapfami} tool enables the development of embedded systems, real-time systems and commercial software through a UML-based visual modeling language or Systems Modeling Language (SysML). GENEXUS\footnote{www.genexus.com/} is a business tool that uses the MDD (Model Driven Development) concept \cite{France2007} to visually model complex systems and has a separate environment for each platform, it also has a language for specifying business rules as an assistance to visual modeling. 


Model-Driven APPLAUSE\footnote{github.com/applause/applause} \cite{dageforde2016generating} is an academic approach that provides a declarative DSL written in XText\footnote{www.eclipse.org/Xtext} with templates for XTend code generation. But the focus of modeling involves technical domain details rather than conceptual elements, maintaining a low level of abstraction. Miravet et al. \cite{miravet2014framework} present another academic approach. The difference between them is the use of XML as a modeling language and the use of the device-independent mobile application generation (DIMAG) to generate native code.

In the work of Inayatullah et al. \cite{inayatullah2019model}, an MDA-based framework (Model-Driven Architecture), provides UML modeling to express the domain concepts, generating hybrid applications and RESTful Services in Asp.Net. APPIAN\footnote{www.appian.com/} and MENDIX\footnote{www.mendix.com} are commercial frameworks that support mobile application development and data interoperability between applications. The difference between them is that the latter also considers the web platform using PhoneGap\footnote{phonegap.com/} as the basis. Interfaces are modeled visually and business rules and behavior are defined at a high abstraction level through custom tool editors, the result is executable code generated. Following the business line, WEBRATIO\footnote{www.webratio.com} \cite{acerbis2015model} is a model-driven development platform based on Interaction Flow Modeling Language\footnote{www.ifml.org/} and with a high abstraction level. WEBRATIO uses the Cordova\footnote{cordova.apache.org/} API as a base and considers the same platforms as MENDIX.



Some work uses MDD to address the development of a single application version. Examples are JUSE4ANDROID \cite{da2014model}, MIMIC \cite{elouali2014mimic} and the research by Cimino and Marcelloni \cite{cimino2012efficient}, which focus exclusively on graphical user interface for the Android platform. Behrens \cite{behrens2010mdsd} presents an environment for textual modeling through a DSL for iOS. Min et al. \cite{min2011uml} and Benouda et al. \cite{benouda2017automatic} use UML to model applications for Windows Phone and generate native code in C\#.

Many approaches try to focus only on Android and iOS, which are currently dominating the market. This is the case of academic approaches such as MD2 \cite{majchrzak2015model}, Chen et al.'s \cite{chen2019automated}, Sabraoui et al.'s \cite{sabraoui2019mdd}, AXIOM \cite{jones2014axiom}, Perchat, Desertot and Lecomte's \cite{perchat2013component}, MOPPET \cite{usman2017product}, Taentzer and Vauper's \cite{taentzer2016model}, Dageförde et al.'s \cite{dageforde2016generating}, Vaupel et al.'s \cite{vaupel2018model}, Rieger et al.'s \cite{rieger2020model}, and industry initiatives such as REACT NATIVE\footnote{facebook.github.io/react-native}, NATIVESCRIPT\footnote{www.nativescript.org}, Flutter\footnote{flutter.dev} and XAMARIN\footnote{www.xamarin.com}. 

The REACT NATIVE framework, developed by Facebook, allows writing code through web technologies (XML, CSS and JavaScript). Its contribution is the generation of readable native code for both platforms (iOS and Android), where it is possible to add new functionality directly in the generated source code. NATIVESCRIPT enables the development of native applications using a single language. In this case, the developer may choose between JavaScript, Angular\footnote{angular.io} or TypeScript\footnote{www.typescriptlang.org} and interface customizations through CSS. The difference between REACT NATIVE and NATIVESCRIPT is that in REACT NATIVE the code is transformed into native language, with the possibility of extension by the developer. And in the second, the code is delivered to devices in JavaScript and interpreted natively through the Virtual Machine (VM) of each platform.

Flutter uses the Dart language to model and generate executable code and is considered an interpreted approach. A compiler converts the code into native language that runs on the device, along with an interface rendering engine. In this way, Flutter does not use native elements for rendering the interface, impairing its performance a little. XAMARIN uses the C\# language, and can obtain between 75\% and 100\% of code reuse between platforms. The applications generated by the tool are native, thanks to the access to the APIs of each platform. However, despite the increase in productivity, the tool does not increase the level of abstraction, keeping the developer involved with technical details.


Chen et al. \cite{chen2019automated} propose a framework to convert user interface (UI) source code from Android to iOS and from iOS to Android. It uses a classification algorithm to identify the components in one platform and then generates corresponding elements in the other platform. Sabraoui et al. \cite{sabraoui2019mdd} use MDD to develop the GUI platform independently. A DSL is used to model the interface, programmed using a grammar written in XText. This DSL can be viewed graphically, as UML models, and through code generators the UI code for the platforms is automatically created.


AXIOM \cite{jones2014axiom} uses UML visual models combined with Abstract Model Tree to represent domain interactions and generic concepts through a Platform Independent Model. Perchat, Desertot and Lecomte \cite{perchat2013component} propose a universal language in which, through a compiler, it is possible to generate native code, but the proposed language has a low abstraction level. MOPPET \cite{usman2017product} is a tool that uses a mixed approach, involving the concepts of Software Product Lines and MDD, defined by the authors as a product-line model-driven engineering approach for native code generation. Case studies have shown reduced effort and increased productivity.

Taentzer and Vauper \cite{taentzer2016model} propose the textual modeling of native applications separated into three submodels: graphical modeling (GUI), data modeling and application behavior (business rules). The main contribution of this work was the generation of native code taking into account information about the context and offline operation.

Dageförde et al. \cite{dageforde2016generating} discuss ways to add the concept of Software Product Lines (SPL) into the MD2 framework. The main contribution of the work was the modularization of the MD2 framework, according to the concepts of SPL, ensuring greater versatility and reuse of modeling artifacts. In the work of Vaupel et al. \cite{vaupel2018model} an approach that uses visual models is presented and specific business rules can be expressed in a textual language, for the generation of native codes (Andoid and iOS). In Rieger et al.'s work \cite{rieger2020model}, the elements for accessibility on mobile devices are included in the framework MD2, aiming to reduce costs in its implementation. 

\begin{table}[htb]
    \centering
    \footnotesize
    \begin{tabular}{|l|c|{c}|{c}|{c}|{c}|{c}|{c}|} \hline
         \textbf{\small Related work} & \textbf{\small C1} & \textbf{\small C2} & \textbf{\small C3} & \textbf{\small C4} & \textbf{\small C5} & \textbf{\small C6} & \textbf{\small C7}  \\ \hline

        {\small HAXE} & {\footnotesize X} & {\footnotesize X}  & {\footnotesize -}  & {\footnotesize -} & {\footnotesize -} & {\footnotesize X} & {\footnotesize X} \\  \hline
        
        {\small KOTLIN} & {\footnotesize X} & {\footnotesize X}  & {\footnotesize -}  & {\footnotesize -} & {\footnotesize -} & {\footnotesize X} & {\footnotesize X}\\  \hline
        
        {\small IBM} & {\footnotesize X} & {\footnotesize X}  & {\footnotesize -}  & {\footnotesize -} & {\footnotesize -} & {\footnotesize X} & {\footnotesize X}\\  \hline        
        
        {\small GENEXUS} & {\footnotesize X} & {\footnotesize -}  & {\footnotesize -}  & {\footnotesize -} & {\footnotesize -} & {\footnotesize X} & {\footnotesize X}\\  \hline
        
        {\small APPLAUSE} & {\footnotesize X} & {\footnotesize X}  & {\footnotesize -}  & {\footnotesize -} & {\footnotesize -} & {\footnotesize X} & {\footnotesize X}\\  \hline 
        
        {\small MIRAVET\cite{miravet2014framework}} & {\footnotesize X} & {\footnotesize X}  & {\footnotesize -}  & {\footnotesize -} & {\footnotesize -} & {\footnotesize X} & {\footnotesize X}\\  \hline  
        
        {\small INAYAT. \cite{inayatullah2019model}} & {\footnotesize X} & {\footnotesize X}  & {\footnotesize -}  & {\footnotesize -} & {\footnotesize -} & {\footnotesize X} & {\footnotesize X}\\  \hline   
        
        {\small APPIAN} & {\footnotesize X} & {\footnotesize X}  & {\footnotesize -}  & {\footnotesize -} & {\footnotesize -} & {\footnotesize X} & {\footnotesize X}\\  \hline 
        
        {\small MENDIX} & {\footnotesize X} & {\footnotesize X}  & {\footnotesize -}  & {\footnotesize -} & {\footnotesize -} & {\footnotesize X} & {\footnotesize X}\\  \hline        
        
        {\small WEBRATIO\cite{acerbis2015model}} & {\footnotesize X} & {\footnotesize X}  & {\footnotesize -}  & {\footnotesize -} & {\footnotesize -} & {\footnotesize X} & {\footnotesize X}\\  \hline   
        
        {\small JUSE4\cite{da2014model}} & {\footnotesize X} & {\footnotesize -}  & {\footnotesize -}  & {\footnotesize -} & {\footnotesize -} & {\footnotesize X} & {\footnotesize X}\\  \hline 
        
        {\small MIMIC\cite{elouali2014mimic}} & {\footnotesize X} & {\footnotesize -}  & {\footnotesize -}  & {\footnotesize -} & {\footnotesize -} & {\footnotesize X} & {\footnotesize X}\\  \hline 
        
        {\small CIMINO\cite{cimino2012efficient}} & {\footnotesize X} & {\footnotesize -}  & {\footnotesize -}  & {\footnotesize -} & {\footnotesize -} & {\footnotesize X} & {\footnotesize X}\\  \hline

        {\small BEHRENS\cite{behrens2010mdsd}} & {\footnotesize X} & {\footnotesize -}  & {\footnotesize -}  & {\footnotesize -} & {\footnotesize -} & {\footnotesize X} & {\footnotesize X}\\  \hline
        
        {\small MIN\cite{min2011uml}} & {\footnotesize X} & {\footnotesize X}  & {\footnotesize -}  & {\footnotesize -} & {\footnotesize -} & {\footnotesize X} & {\footnotesize X}\\  \hline      
        
        {\small BENOUDA\cite{benouda2017automatic}} & {\footnotesize X} & {\footnotesize -}  & {\footnotesize -}  & {\footnotesize -} & {\footnotesize -} & {\footnotesize X} & {\footnotesize X}\\  \hline

        {\small MD2\cite{majchrzak2015model}} & {\footnotesize X} & {\footnotesize X}  & {\footnotesize -}  & {\footnotesize -} & {\footnotesize -} & {\footnotesize X} & {\footnotesize X}\\  \hline      
        
        {\small CHEN\cite{chen2019automated}} & {\footnotesize X} & {\footnotesize X}  & {\footnotesize -}  & {\footnotesize -} & {\footnotesize -} & {\footnotesize X} & {\footnotesize X}\\  \hline
        
        {\small SABRAOUI\cite{sabraoui2019mdd}} & {\footnotesize X} & {\footnotesize X}  & {\footnotesize -}  & {\footnotesize -} & {\footnotesize -} & {\footnotesize X} & {\footnotesize X}\\  \hline     
        
        {\small AXIOM\cite{jones2014axiom}} & {\footnotesize X} & {\footnotesize -}  & {\footnotesize -}  & {\footnotesize -} & {\footnotesize -} & {\footnotesize X} & {\footnotesize X}\\  \hline      
        
        {\small PERCHAT\cite{perchat2013component}} & {\footnotesize X} & {\footnotesize X}  & {\footnotesize -}  & {\footnotesize -} & {\footnotesize -} & {\footnotesize X} & {\footnotesize X}\\  \hline

        {\small MOPPET\cite{usman2017product}} & {\footnotesize X} & {\footnotesize -}  & {\footnotesize -}  & {\footnotesize -} & {\footnotesize -} & {\footnotesize X} & {\footnotesize X}\\  \hline  
        
        {\small TAENTZER\cite{taentzer2016model}} & {\footnotesize X} & {\footnotesize -}  & {\footnotesize -}  & {\footnotesize -} & {\footnotesize -} & {\footnotesize X} & {\footnotesize X}\\  \hline       
        
        {\small DAGEFÖRDE\cite{dageforde2016generating}} & {\footnotesize X} & {\footnotesize X}  & {\footnotesize -}  & {\footnotesize -} & {\footnotesize -} & {\footnotesize X} & {\footnotesize X}\\  \hline  
        
        {\small VAUPEL\cite{vaupel2018model}} & {\footnotesize X} & {\footnotesize X}  & {\footnotesize -}  & {\footnotesize -} & {\footnotesize -} & {\footnotesize X} & {\footnotesize X}\\  \hline    
        
        {\small RIEGER\cite{rieger2020model}} & {\footnotesize X} & {\footnotesize X}  & {\footnotesize -}  & {\footnotesize -} & {\footnotesize -} & {\footnotesize X} & {\footnotesize X}\\  \hline
        
        {\small REACT \newline NATIVE} & {\footnotesize X} & {\footnotesize X}  & {\footnotesize -}  & {\footnotesize -} & {\footnotesize -} & {\footnotesize X} & {\footnotesize X}\\  \hline 
        
        {\small NATIVESCRIPT} & {\footnotesize X} & {\footnotesize X}  & {\footnotesize -}  & {\footnotesize -} & {\footnotesize -} & {\footnotesize X} & {\footnotesize X}\\  \hline  
        
        {\small FLUTTER} & {\footnotesize X} & {\footnotesize X}  & {\footnotesize -}  & {\footnotesize -} & {\footnotesize X} & {\footnotesize X} & {\footnotesize X}\\  \hline  
        
        {\small XAMARIN} & {\footnotesize X} & {\footnotesize X}  & {\footnotesize -}  & {\footnotesize -} & {\footnotesize -} & {\footnotesize X} & {\footnotesize X}\\  \hline  
        
        {\small THIS WORK} & {\footnotesize X} & {\footnotesize X}  & {\footnotesize X}  & {\footnotesize X} & {\footnotesize X} & {\footnotesize X} & {\footnotesize X}\\  \hline

    \end{tabular}
    \caption{Brief comparison between related work and this one, in terms of the identified contributions}
    \label{tab:ComparativoAbordagens}
\end{table}

In summary, both academia and industry are successful in delivering a single way to develop software for multiple platforms, allowing the developer to reuse code across platforms and domains more efficiently, helping to reduce development and maintenance costs. But they have strong ties with the supported platforms, not giving the developer the choice to include new platforms or extend existing support without considerable effort. They also normally consider that all platforms will receive the same functions, not allowing different devices to receive different functions. We formalize all these desired contributions as follows:


\begin{description}
\item[C1.] \textbf{High abstraction level modeling in a single environment.} The software engineer should be able to specify system concepts at an abstraction level that sits above platform-specific details.
\item[C2.] \textbf{Reuse of similar concepts between platforms through modeling.} Many concepts, such as persistent entities and functions, have to be repeatedly created, with small differences, for each platform. The approach must allow these to be reused in all platforms.
\item[C3.] \textbf{Reduced system configuration costs.} This includes distributing functionality among platforms and switching the platforms being used in the system.
\item[C4.] \textbf{Inclusion of new platforms with a small impact.} The ability to include new platforms into the system, even if they were not initially supported.
\item[C5.] \textbf{Extension of the approach through platform models.} Existing platform support should be extensible to better suit the needs of a particular system.
\item[C6.] \textbf{Code reuse across domains.} Cross-domain concepts and functions should be specified in such a way that they can be easily reused in any domain.
\item[C7.] \textbf{Development and maintenance cost reduction.} The approach's support for abstraction, reuse and code generation should bring the potential to reduce costs during the development life cycle. 
\end{description}

Table \ref{tab:ComparativoAbordagens} presents a simple comparison between existing work and our proposed approach, in terms of these seven contributions. The table is just a summary of the essence of each solution. There may be specific ways to achieve these contributions that were not accounted for because this is not explicitly mentioned in the research paper or documentation. All solutions deliver higher abstraction level development (\textbf{C1}), cross-domain reuse (\textbf{C6}) and cost reduction (\textbf{C7}). But none have been designed to allow functionality distribution (\textbf{C3}) and the inclusion of new platforms (\textbf{C4}). Platform extension (\textbf{C5}) is a recent addition to Flutter in the form of ``platform channels''\footnote{flutter.dev/docs/development/platform-integration/platform-channels}, which are somehow similar to this approach's concept of global functions, as discussed later.

\section{A holistic approach for cross-platform software development}
\label{sec:approach}

\begin{figure*}[htb]
\begin{center}
\includegraphics[width=0.9\linewidth]{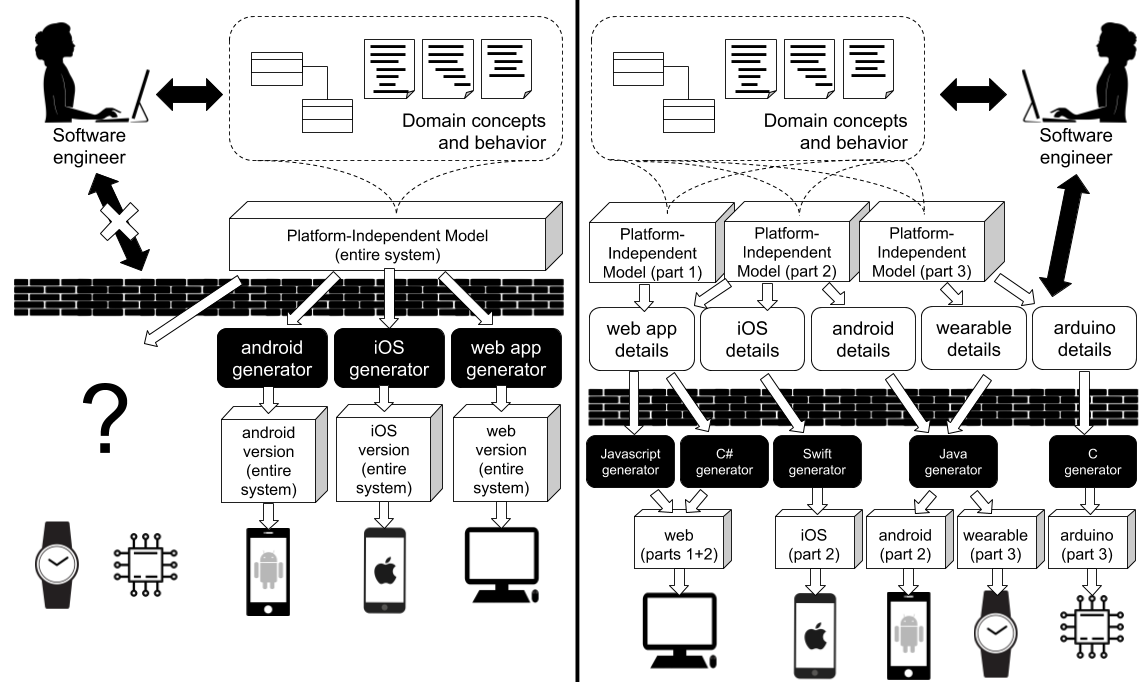}
\caption{An overview of a typical MDD cross-platform solution (left) compared with the approach proposed in this research (right).}
\label{fig:overviewAndComparison}
\end{center}
\end{figure*}

In order to overcome the limitations of existing work, this research proposes a holistic view of cross-platform software development\footnote{https://github.com/JulianoZ/CrossPlatformApproach/}. For that, it borrows the main idea from Model Driven Development (MDD) \cite{France2007}, which is to combine generative programming, software transformation techniques and domain-specific languages (DSL). But just using MDD for cross-platform development has already been proposed and successfully tested by others, as discussed in the previous section. The main difference introduced in this research is how platform-specific details are taken into consideration, as shown in Figure \ref{fig:overviewAndComparison}.

The left side of Figure \ref{fig:overviewAndComparison} shows a typical MDD-based cross-platform solution. The software engineer works on a higher abstraction level, to produce a platform-independent model. She is shielded from the details of the underlying platforms by means of platform-specific generators and transformations. These artifacts are provided to the developer as black-box components, to be used as they are or with some parameterization. As a result, the developer can focus on business domain concepts, as intended by MDD. But there is little flexibility to modify platform details or include new platforms. The only solution would be to modify the code generators or transformations, but since these were designed to be used as black-box components, this is a difficult task. Also, code generators produce different versions of the same software. This means that all functions for the entire system are generated for all platforms, equally.

In contrast, this approach (right side of Figure \ref{fig:overviewAndComparison}) takes a more holistic viewpoint. Business domain concepts and the platform details are meant to be under the control of the developer. The only black-box components are the language-specific code generators, which will only need to change if the target language changes, what is not likely to occur very often. This allows platforms to be modified or added by the developer. The approach also provides the ability to choose where each part of the software will be deployed. For example, an administrative area of a system may be targeted at the web only. The main front-end may be targeted at the web and mobile platforms, and a part of a system dealing with sensors might be targeted at wearable and arduino platforms. All these specifications are platform-independent, and the entire solution is treated as a single software entity.

\begin{figure}[htb]
\begin{center}
\includegraphics[width=0.9\linewidth]{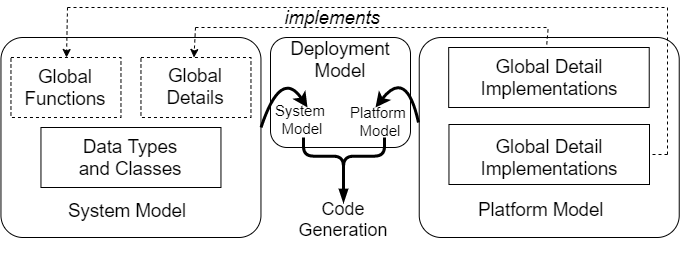}
\caption{Cross-platform approach elements.}
\label{fig:approachElements}
\end{center}
\end{figure}

This is made possible with a Generic Purpose Language (GPL) designed for this goal. This GPL is a textual programming language with some high-level constructs for business domain concepts, detailed behavior, platform details and functionality distribution. So, if a different way of generating code for a specific platform is needed, or if a new platform needs to be supported, the developer can use the GPL instead of dealing with code generators or modifying generated code. Figure \ref{fig:approachElements} shows the three models that need to be specified by the software engineer when creating a cross-platform system using the GPL: the system model (left), the platform model (right) and the deployment model (middle).

\subsection{The system model}

The system model is composed of three main parts: Data Types and Classes, Global Functions and Global Details.

The first part of a system model are \textbf{Data Types and Classes}. These are regular object-oriented constructs to specify a system's structure and behavior. Listing \ref{lst:dataTypesAndClasses} shows an example of this part of the system model, where three data types (lines 1-3) and one class (lines 4-12) are defined. The class has two attributes (line 5) and one operation (lines 6-12). Currently, the GPL has support for most object-oriented programming language features, however it is still a prototype and may have some missing implementation details.

\begin{lstlisting}[caption={Data type and class declarations using the GPL.},label={lst:dataTypesAndClasses}]
datatype int
datatype string
datatype double
class Coordinates {
   double x, y;
   operation dist(o: Coordinates): double {
      dx: float
      dy: float
      dx := o.x - x
      dy := o.y - y
      return sqrt(dx^2 + dy^2)
} }
\end{lstlisting}

The second part of a system model are \textbf{Global Functions}. These are the most important elements of the approach, as they establish the relationship between platform-independent and platform-specific models. Consider for example a function for obtaining the device location. Each platform (android, iOS, web, etc.) will have a different implementation, possibly with a different function signature. But, in essence, they could be represented as a function, called ``getLocation()'' that returns a pair of coordinates (latitude and longitude). In the approach, this is exactly what a global function is: an abstraction for a function that is needed by the system, has the same (or a very similar) signature, but requires platform-specific implementations.

Listing \ref{lst:globalFunctions} shows an example of global function declarations. It is possible to see that the approach supports cross-domain global functions, such as \verb|getGPSPosition| (line 1), to obtain a device's position, for example. This type of function can be used in different applications, from different domains. The approach also supports domain-specific global functions, such as \verb|InsertProductIntoCart|, \verb|SelectCustomer| and \verb|SelectProduct| (lines 2-4). These are specific to an e-commerce domain.

\begin{lstlisting}[caption={Example of Global Function Declarations.},label={lst:globalFunctions}]
global getGPSPosition(): Coordinates
global SelectCustomer(id: string): Customer
global SelectProduct(id: string): Product
global InsertProductIntoCart(customer: Customer, product: Product, quantity: int): string
\end{lstlisting}

A global function is supposed to be supported by different platforms. This is possible as long as the platforms are able to implement the declared function prototype. For example, most smartphones are capable of obtaining their positions from GPS satellites, therefore they will probably be able to provide a compatible implementation for function \verb|getGPSPosition|. But a desktop or portable computer does not have a GPS receiver, therefore it will not be able to implement this function\footnote{Computers have location systems, but they normally depend on antennas and might not function in a farm or in the ocean, for example.}. As discussed before, the approach does not assume that all platforms will run identical versions of the entire system, therefore not all global functions will be implemented by all platforms. Instead, the approach assumes that the system requires that these functions will be implemented by at least one platform, otherwise the system will not be able to execute properly.

The third part of a system model are \textbf{Global Details}. These represent platform-specific code that only makes sense in a particular platform and that cannot be generalized as a global function. In other words, any piece of code that is not the implementation of a global function, such as additional variables, methods and annotations, that must somehow appear in the generated code, must be specified as global details. For example, global function \verb|SelectCustomer| from Listing \ref{lst:globalFunctions}, when deployed in the web platform, requires a database connection to be available. The implementation for this function could contain code that creates a new connection every time the function is called, but a more efficient approach would be to create the connection once, store it as a class attribute, and reuse it across calls. A class attribute would fall outside the implementation of a global function, and this is where global details are useful. They can be used to include class attributes, additional methods (other than the global function itself), annotations, among other code.

Global details consist of a declaration, implementation and use. The declaration is very simple, as shows Listing \ref{lst:globalDetails}. 
The implementation of this example, which will create a connection attribute and helper methods, will be described later (Listing \ref{lst:globalDetailsImpl}).

\begin{lstlisting}[caption={Example of Global Details Declarations.},label={lst:globalDetails}]
globalDetails DBConnection 
\end{lstlisting}

Global functions and details can be used by any class. An example is shown in Listing \ref{lst:ShoppingCartDAO}. The use of a global details in a class is declared with the keyword \verb|usesGlobalDetails| (line 2). In this example, these details represent database connection initialization code. The use of a global function in a class is declared with the keyword \verb|usesGlobal| (lines 3-5). To improve readability, the complete global function declaration has to be repeated into the class. Then, these functions can be invoked normally (lines 11-13) as if they were local functions, except for the keyword \verb|global|.

\begin{lstlisting}[caption={A class using global functions and details.},label={lst:ShoppingCartDAO}]
class ShoppingCartDAO {		
    usesGlobalDetails DBConnection
    usesGlobal SelectCustomer(id: string): Customer
    userGlobal SelectProduct(id: string): Product
    usesGlobal InsertProductIntoCart(customer: Customer, product: Product, quantity: int): string	
    operation addProduct(custId: string, prodId: string, quantity: int): list<string> {
        objCust: Customer
        objProd: Product
        newProd: string
        accessData: list<string>		
        objCust := global SelectCustomer(custId)
        objProd := global SelectProduct(prodId)
        newProd := global InsertProductIntoCart(objCust, objProd, quantity)
}	}
\end{lstlisting}

Each global function will have a different implementation on each platform. In the example of Listing \ref{lst:ShoppingCartDAO}, the global invocations in lines 11-13 might refer to a SQL database if the class is deployed into a web platform, or to a SQLite database if the class is deployed into a mobile platform. But here in the system model, this is intentionally left undefined. It shouldn't matter which platform will run this code, it only needs a compatible implementation for these functions.

Global functions and global details are very similar to the concept of platform channels present in Flutter. There are two main differences: first, in Flutter only the predefined platforms are supported, while in this approach any platform can implement any global function; second, in Flutter, channels are linked to the implementation during runtime, while here they are included in the generated code, thus being linked during compilation time. Flutter's runtime-based approach makes it easier to reuse channels as black-box components, in different apps. This approach's generation-based process, together with global details, makes it easier to extend or customize platform-specific code.

\subsection{The platform model}
\label{sec:platformModel}

The platform model consists of implementations for the global functions and global details, and is defined for a particular language. The global function prototypes are declared using the GPL syntax, but the actual implementation of these functions may be written using either the GPL or the platform's native programming language. Using the platform language is preferred, as it allows more direct access to all of the platform's features. Listing \ref{lst:platform} shows three example of platforms: Web using C\# (lines 1-5), iOS using Swift (lines 6-10) and Augmented Reality using C\# (lines 11-15), an entirely new platform developed for this project. 

\begin{lstlisting}[caption={Implementation of global functions.},label={lst:platform}]
platform Web: CSharp {	
  implementsGlobal InsertProductIntoCart(c: Customer, p: Product, quantity: int): string {
    //C# code that implements the function on the web
    //platform using SQL for persistence
}  }
platform iOS: Swift{	
  implementsGlobal InsertProductIntoCart(c: Customer, p: Product, quantity: int): string {
    // Swift code that implements the function on iOS
    // using SQLite for device persistence
}  }  
platform AugmentedReality: CSharp {
   implementsGlobal GetDiskImagesAsync(picturesFolder: StorageFolder): list<img> {
     //C# code that implements the function on  
     //the platform for Augmented Reality
}  }
\end{lstlisting}

As discussed before, the holistic approach does not assume that all platforms will run identical versions of the entire system. In the example of Listing \ref{lst:platform}, it makes no sense to provide an implementation for the \verb|InsertProductIntoCart| global function for the Augmented Reality platform, as this platform does not have data persistence support. Therefore, only the Web and iOS platforms will contain an implementation for this global function. The Augmented Reality platform will have an implementation for global function \verb|GetDiskImagesAsync|, for displaying images.

The software engineer has flexibility to provide custom implementations for each platform. In the example of Listing \ref{lst:platform}, each platform adopts a different persistence mechanism: SQL-based, for Web/C\# platform (lines 3-4), and SQLite-based, for the iOS/Swift platform (lines 8-9).

Also to increase the flexibility, the approach has support for Apache Velocity templates\footnote{\url{velocity.apache.org}}, making implementation code more flexible and increasing their reusability. As an example of this additional feature, Listing \ref{lst:InsertObject} shows a domain-independent way to declare, implement and use a global function \verb|SelectObject|, which can query any object from a SQL database. In the declaration part (line 2), there is nothing different from a normal global declaration, except for the generic type \verb|<E>|, which can be specified later.


In the implementation (lines 4-21), triple apostrophes (\verb|'''| in lines 5 and 21) are used to delimitate the template contents. Apache Velocity tags (starting with \# and \$) are used to query the generic type \verb|<E>| in search for its features. It is possible, for example, to use \verb|E|'s name (\verb|$E.name| in lines 6 and 12) to specify the query parameters and specific SQL values (line 12), or to query \verb|E|'s attributes (\verb|$E.attributes| in line 12) to generate an SQL statement.

When using this global function, it is necessary to specify a concrete class to be used for code generation. In this example, class \verb|Order| (line 23) will be the object being inserted. The approach takes care of linking these definitions and generating correct code for the platform. The result is that this global function may be reused in different domains.

\begin{lstlisting}[caption={Declaring, Implementing and Using a domain-independent global function using Apache Velocity.},label={lst:InsertObject}]
//Declaration
global <E> SelectObject(ord: string): list<E> 
//Implementation
implementsGlobal <E> SelectObject(ord:string):list<E>{
'''     
   List<$E.name> listObj = new List<$E.name>(); 
   $E.name obj = new $E.name();
   Connection objCon = new Connection();
   MySqlConnection Conn = new MySqlConnection();
   Conn = objCon.OpenConnection();
   MySqlCommand command = Conn.CreateCommand();
   command.CommandText = "select #foreach($f in $E.attributes)${f.name}#if($foreach.hasNext) , #end#end from $E.name " + ord;
   for(Reader.Read()){
      #set($count = 0)	
      #foreach($f in $E.attributes)
         obj.set${f.name.substring(0,1).toUpperCase()}${f.name.substring(1)}(Reader.get${f.type.name.substring(0,1).toUpperCase()}${f.type.name.substring(1)}($count));
         #set($count = $count + 1)                 
      #end                 
      listObj.add(obj)
   } return listObj;	
''' }
// Use
usesGlobal <Order> SelectObject(ord: string): list<Order>

 \end{lstlisting}

In addition to global functions, the platform model also contains implementations of global details. Listing \ref{lst:globalDetailsImpl} provides an example of an implementation of global details. As discussed before, these are used to create platform-specific code (fields and methods) that cannot be generalized as global functions. In this case, connecting to MySQL web platform in C\# requires a class attribute for the connection (line 4), a method for opening the connection (lines 5-9) and a method for closing the connection (lines 10-12). These will be used internally by global function implementations.

\begin{lstlisting}[caption={implementation of global details for a MySQL database connection in a web platform.}, label={lst:globalDetailsImpl}]
platform Web : CSharp {
    implementsGlobalDetails DBConnection{
    '''
        private MySqlConnection Conn;
        public MySqlConnection OpenConnection(){
            Conn = new MySqlConnection("server=127.0.0.1;database=ecommerce;uid=18feb24ac5h;pwd=pass");
            Conn.Open();
            return Conn;   
        }
        public void CloseConnection(){
            Conn.Close();
        }    	
    '''	
}
\end{lstlisting}

If a platform has no need for specific details, it may leave the implementation for the global details empty. As a result, no additional code will be generated for that platform.

\subsection{The deployment model}

The deployment model simply defines which classes will execute on which platforms, serving as a guide for code generation. The snippet shown in Listing \ref{lst:deploy} presents an example of deploying different classes on different platforms. The only restriction is that a platform chosen for deployment of a particular class must have an implementation for all global functions and details used by that class. In this example, the \verb|User| and \verb|UserDAO| classes are being deployed to all platforms. The \verb|ShoppingCartDAO| class is being deployed on Android (line 8) and iOS (line 9), but not on the web.

\begin{lstlisting}[caption={Deployment Configuration (deploy).},label={lst:deploy}]
deploy Ecommerce {
    User: Android
    User: iOS
    User: Web    
    UserDAO: Android
    UserDAO: iOS
    UserDAO: Web
    ShoppingCartDAO: Android
    ShoppingCartDAO: iOS
}
\end{lstlisting}

\subsection{Code generation}

As discussed before, the only black-box components are the language-specific code generators. Currently, there are generators for the Java, C\#, and Swift languages, so it is possible to define platforms based on these languages. New generators can be added to the approach by implementing a mapping between the GPL constructs and target language constructs, using Eclipse Xtext\footnote{\url{www.eclipse.org/Xtext/}} and Apache Velocity\footnote{\url{velocity.apache.org}}.

\begin{figure*}[htb]
    \centering
    \includegraphics[width=0.9\linewidth]{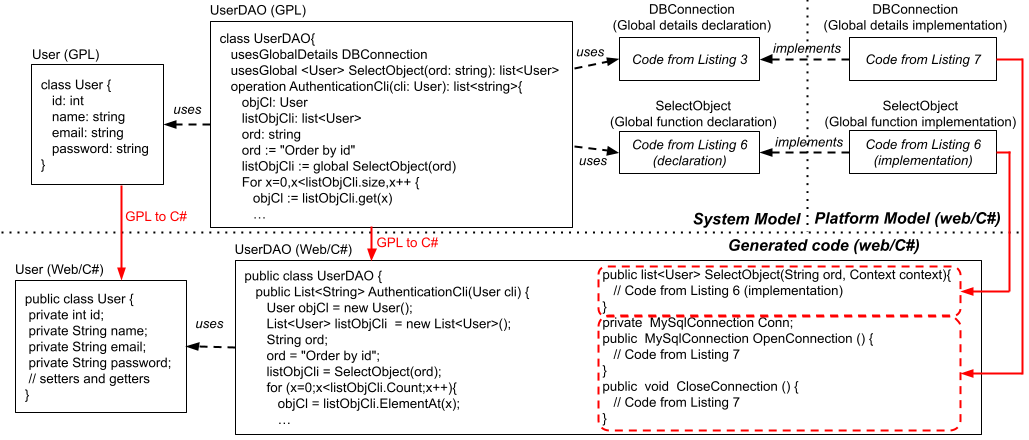}
    \caption{Code generation result for classes User and UserDAO in the web/C\# platform}
    \label{fig:codeGenerationResult}
\end{figure*}

Figure \ref{fig:codeGenerationResult} shows an example of deploying classes \verb|User| and \verb|UserDAO| into the web/C\# platform. It is possible to see how the system and platform models from the examples shown in this section are integrated into the generated code, which will contain all that is necessary for proper execution.

\section{Evaluation}
\label{sec:Evaluation}

Four evaluations were conducted with the goal of finding evidence regarding the seven main contributions of the approach presented in the end of Section \ref{sec:relatedWork}. The \textbf{first} evaluation was a proof-of-concept, focused on verifying if the approach can be used to produce a real cross-platform system. The \textbf{second} evaluation was an expert evaluation. It was conducted with five experts and focused on gathering specialized opinions regarding the expected contributions. The \textbf{third} evaluation was similar to the second, but with novice developers. The idea was to test if the approach can shield these developers from the technical details, allowing them to develop software for these platforms even without the necessary knowledge. Finally, the \textbf{fourth} evaluation involved using the approach to create test cases for different platforms, following observations made by the experts.

\subsection{Proof-of-concept}
\label{sec:proofOfConcept}

This evaluation was conducted by the researchers, and had three stages. The \textbf{first} stage occurred before the approach was developed, and consisted in the development of a complete cross-platform system for the e-commerce domain. Since the approach did not exist then, the development followed an ad-hoc process. Three platforms were supported: web, android and iOS. The system's architecture follows the Model View Controller (MVC) \cite{bucanek2009model} style,. The system has basic CRUD (Create, Retrieve, Update and Delete) functions, an area for customers to browse and order products using a virtual shopping cart, and an administrative area for sellers to process the orders. The customer area also includes functions to detect the customer location. Mobile platforms (Android and iOS) only have the customer area, while web platforms have both the customer and administrative areas.

The \textbf{second} stage occurred after the approach was developed, and consisted in migrating the existing system into the approach, i.e. creating the system, platform and deployment models. The main challenge in this stage was to decide which parts of the system would become system models (data types and classes) and which would become platform models (global functions and global details). As discussed before, the main idea is to try to generalize common functions with similar signatures as global functions.

Device location support was obviously part of the platform models, so it was defined as global function and details. For the other parts, we identified the similarities between the platforms in the original code. We observed that around 79\% of the developed system consists in concepts and functions that are related either to the MVC architectural style or to CRUD functions. Therefore, we defined global functions and details for these concepts, and used Velocity templates to implement them. As a result, two platforms for each device type were defined: one for the ``Model'' layer and one for the ``Controller'' layer. For this proof-of-concept, we did not migrate the ``View'' layer into the approach, as it involves visual design and layout definition that is better performed directly in the platform's native IDE.


In the \textbf{third} stage, a fourth platform was included, to test the approach's flexibility to include new platforms. To really put this flexibility to the test, a new platform was created solely for this research, so that it would be demonstrated that the approach does support unforeseen devices. The platform consisted of an augmented reality (AR) device that can project images on real-world objects, allowing customers to see products on a display being highlighted by the device.

\begin{figure}[t]
\begin{center}
\includegraphics[width=0.8\linewidth]{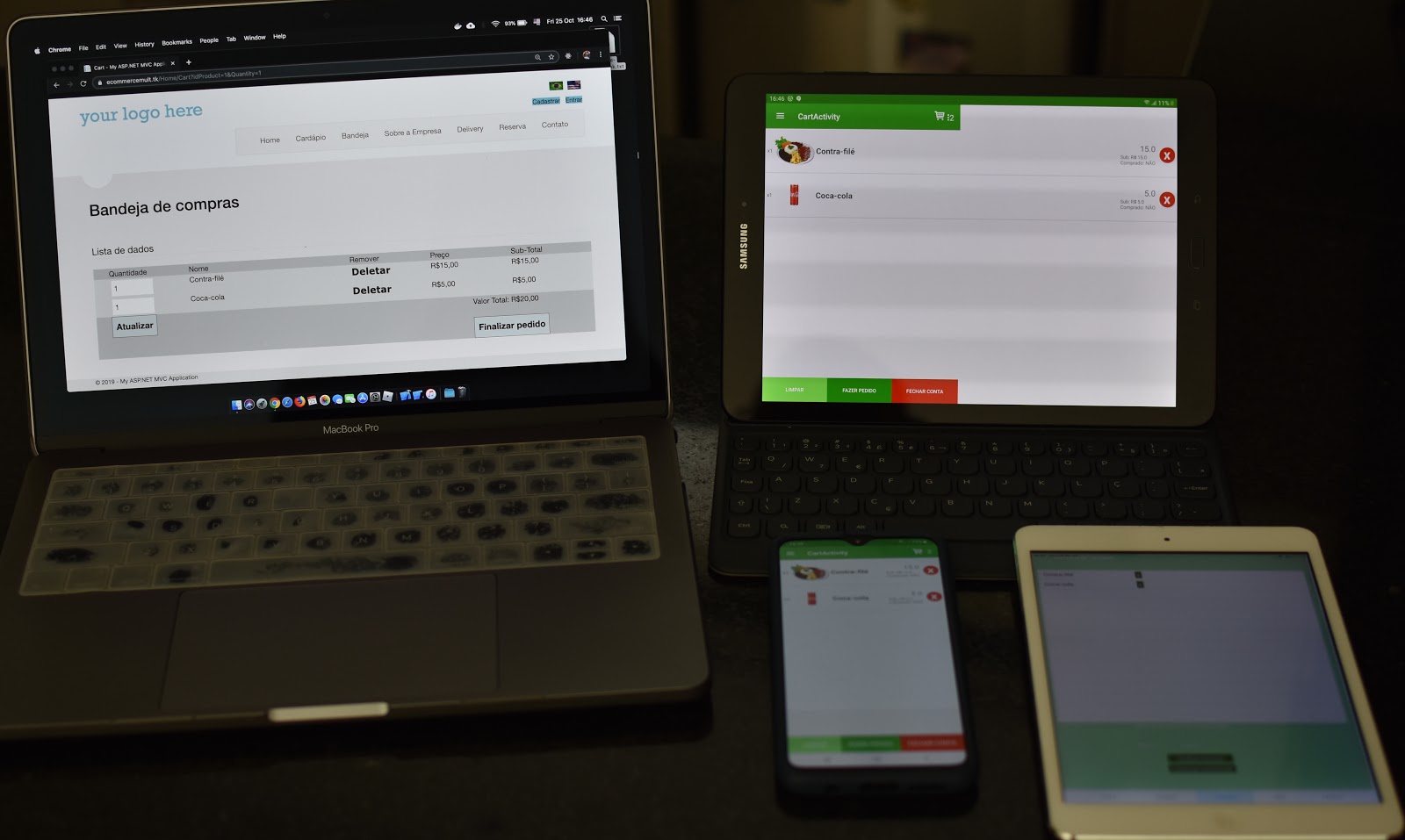}
\caption{Shopping cart screen of the e-commerce application shown on four real devices, in three different platforms}
\label{fig:multiplataforma}
\end{center}
\end{figure}

\begin{figure}[t]
\begin{center}
\includegraphics[width=0.8\linewidth]{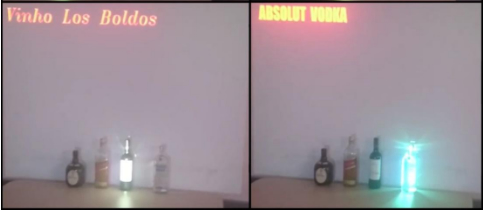}
\caption{Augmented reality device highlighting real-world products and projecting their names above as they are selected in the e-commerce application}
\label{fig:ar}
\end{center}
\end{figure}

Figures \ref{fig:multiplataforma} and \ref{fig:ar} show five different devices running the application. In Figure \ref{fig:multiplataforma} four devices are showing the shopping cart screen: a desktop computer for the web platform, two tablets (Android and iOS), and one smartphone (Android). Figure \ref{fig:ar} shows the augmented reality device projecting images over some products on display, to highlight them as they are selected by the customer.

\subsubsection{Results}
\label{sec:results}

With the proof-of-concept developed and running, the seven contributions of the approach were evaluated.

Table \ref{tab:locAnalysisProofOfConcept} shows a comparison in terms of LOC (lines of code), without and with the approach, excluding the ``view'' layer of the MVC architecture as it was not implemented in the proof-of-concept. The three zeroes in the table are explained as follows: 1) there is no platform-specific code for the augmented reality (AR) device without the approach, as it was only included in the second stage; 2) there is no platform-independent code without the approach, as it is written entirely in the platform's language and can not be reused in the other platforms (Android, iOS and Web platforms use different languages in the proof-of-concept); and 3) there is also no platform-independent code written in a native language with the approach.

When comparing the total size of the two versions of the system (last line of Table \ref{tab:locAnalysisProofOfConcept}), there is a large reduction in terms of LOC. This is expected, as the generic CRUD functions were made generic and were reused for all persistent classes. This observation indicates that contribution \textbf{C7} (Development and maintenance cost reduction) is being delivered by the approach. The LOC analysis also indicates that the reuse of similar concepts between platforms through modeling is made possible with the approach (contribution \textbf{C2}). Each class was deployed in two or three platforms. This also contributed to the reduction in LOC.

\begin{table}[b]
\scriptsize
    \centering
    \begin{tabular}{cc|c|cc}
         
& & Without & \multicolumn{2}{c}{With approach} \\
& & approach & GPL & Native \\ \hline
Platform- & Android & 1204 & 16 & 207 \\ \cline{2-5}
specific & iOS & 1503 & 16 & 157 \\ \cline{2-5}
code & Web & 1012 & 10 & 97 \\ \cline{2-5}
& AR & 0 & 55 & 89 \\ \hline
\multicolumn{2}{c|}{Platform-independent code} & 0 & 238 & 0 \\ \hline
\multirow{2}{*}{Total} & & \multirow{2}{*}{3719} & 335 & 550 \\ \cline{4-5}
& & & \multicolumn{2}{c}{885} \\ \hline
    \end{tabular}
    \caption{LOC analysis for the proof-of-concept}
    \label{tab:locAnalysisProofOfConcept}
\end{table}

Finally, the LOC analysis indicates that the approach leads to a better focus on conceptual work. While all 3719 lines of code of the original system mix generic domain concepts and platform details, with the approach there are 238 lines of code that correspond exclusively to high-level concepts. The remaining lines of code, created using the GPL and native code, focuses on platform details alone. This separation helps to raise the abstraction level in which the software engineer can work (contribution \textbf{C1}).

The proof-of-concept have also shown that distributing functionality across the platforms and devices is possible and requires little effort (contribution \textbf{C3}). It was very simple to replicate the original system deployment with the approach, with the added benefit of avoiding code duplication.

The third stage of the proof-of-concept demonstrated the ability to include new platforms (contribution \textbf{C4}). A newly created device for augmented reality was included in the system without requiring an entirely new development. Existing classes were targeted at the AR platform during deployment and new global functions were successfully added.

In summary, the development of the proof-of-concept led to indications that all contributions, except \textbf{C5} (extension of the approach through platform models) and \textbf{C6} (code reuse across domains), are being achieved with the approach.

\subsection{Expert Evaluation}
\label{sec:SecondEvaluation}

A second evaluation consisted in an experiment involving experts with large experience in web and mobile technologies (minimum 10 years) and some experience with multiplatform frameworks (desirable but not mandatory). Using convenience sampling, the researchers searched through their contacts in academia and industry in search for volunteers. In the end, five experts that fulfill the requirements were selected. All of them have experience in different programming languages, system architectures, programming languages and databases. Two experts (\textbf{E1} and \textbf{E2}) work in software companies, and have large software development experience. Two experts have academic and research experience, one being a PhD candidate in computer science (\textbf{E3}) and one being a PhD teacher in a computer technology center (\textbf{E4}). The fifth expert (\textbf{E5}) has a PhD in computer science and has experience in the industry but is currently a teacher and researcher in a technology center.

In this evaluation, the experts had to perform five different software development tasks using the approach. The tasks were defined as follows:

\begin{description}
\item[Task 1:] implement a shopping cart for a restaurant application for three platforms (Android, iOS and Web). 

\item[Task 2:] reconfigure how each part of the proof-of-concept system is deployed in the different platforms. The experts were also asked to use the GPS location in some specific platforms.

\item[Task 3:] include the support for a new database into the proof-of-concept.

\item[Task 4:] create a new project, for a different domain (a library), with functions for user authentication and borrowing books.

\item[Task 5:] include a new platform (Android smartwatch) in the e-commerce proof-of-concept, where the product listing functionality would be deployed.

\end{description}

In all tasks, experts could reuse all models and code from the e-commerce proof-of-concept.

The evaluation with experts had three stages. In the first stage, the experts had to learn how to use the approach and how the proof-of-concept system is organized. A repository containing documents and tutorials\footnote{The complete experts material are available at: \url{https://github.com/JulianoZ/CrossPlatformApproach}} was made available for the experts. The repository also has the source code for the proof-of-concept system, including all models (system, platform and deployment). The code was fully documented to facilitate understanding. Experts were asked to record the time spent in the first stage.

In the second stage, the experts had to perform the tasks. They used their personal computers, using their own free time. They were also free to contact the researchers to ask questions. The time spent in each task should be recorded.

In the third and final stage, each expert was interviewed. A script with eleven questions (\textbf{Q1}-\textbf{Q11}) was defined:

\begin{description}

\item[Q1.] What is your opinion regarding the learning curve of the approach?
\item[Q2.] What is your opinion regarding the programming language (GPL) used in the approach? Is the language easy or difficult to use? Is it intuitive? How complex is the language?
\item[Q3.] What is your opinion regarding the platform models provided by the approach? Is it possible to include all technical details through global functions? Is it possible to reuse technical details in systems from different domains? Is it possible to extend the approach through platform models?
\item[Q4.] What did you think of the global functions and generic parameters, used together with Velocity templates, to create dynamic functions in the platform models? Can they automate code generation? Do they allow the creation of dynamic functions during coding? Do they have the potential to create custom code for different entities?
\item[Q5.] What did you think of the deployment model? Comment on: the distribution of functionality, switching platforms for some functions, such as receiving orders in the examples, and managing code generation through this model.
\item[Q6.] Do you believe the approach can reduce cross-platform development costs? \item[Q7.] How do you evaluate the maintenance of a cross-platform system developed with the approach? Were you able to change the modeling (GPL) and reuse the generated code? For example, switching a variable from float to double or change the way the shopping cart stores information (in a database or in memory).
\item[Q8.] How do you evaluate the benefits provided by the approach in the development of cross-platform systems?
\item[Q9.] Please rate your experience with the approach in a scale from 1 to 10, where 1 means a very poor experience and 10 means a very good experience.
\item[Q10.] Do you have another idea to reduce costs in cross-platform development?
\item[Q11.] Please rate each of the following contributions in a scale from 1 to 10. (all contributions, except C7, are listed for the expert to rate)

\end{description}

Most questions were open-ended. Therefore, in order to analyze the responses, the interviews were recorded and a transcript was written to facilitate analysis\footnote{The complete experts transcripts are available at:  \url{https://github.com/JulianoZ/CrossPlatformApproach}}. The transcripts were analyzed separately by the two authors of this paper. During this analysis, the researchers annotated those parts that indicate some evidence in favor (\textbf{C+}) or against (\textbf{C-}) one or more of the contributions. The two sets of annotations were then compared and a common agreement was reached after discussions. Finally, the number of annotations was counted in order to provide a quantitative view of the collected evidence. In this process, when there were more than one annotation of the same type in the same response from the same expert, only one was accounted for. This was necessary because sometimes the experts were repeating themselves during their responses to a question.

\subsubsection{Quantitative analysis}

All tasks were completed successfully by all experts. Table \ref{tab:timeAnalysis} summarizes the time spent during learning and during each task, for each expert. As it can be seen, the average time spent by experts during the entire study was 14h10. Learning took, in average, 6h24, with the remaining time spent in the five tasks. There was some variability from expert to expert, regarding some tasks. Expert \textbf{E3} was the fastest in most tasks, and task \textbf{T2} was the longest, followed by \textbf{T5}. Despite these differences, we consider that the experts were not significantly different from each other and from what was expected, indicating that the tasks were performed more or less in the same way by the experts.

\begin{table}[htb]
    \scriptsize
    \centering
    \begin{tabular}{|c|c|c|c|c|c|c|c|c|} \hline
         \textbf{Exp.} & \textbf{Learn.} & \textbf{T1} & \textbf{T2} & \textbf{T3} & \textbf{T4} & \textbf{T5} & \textbf{Total} \\ \hline
         \textbf{E1} & 6h00 & 1h00 & 2h20 & 0h40 & 0h45 & 1h30 & \textbf{12h15} \\ \hline
         \textbf{E2} & 4h00 & 0h40 & 3h40 & 0h45 & 0h30 & 4h00 & \textbf{13h35} \\ \hline
         \textbf{E3} & 7h00 & 0h25 & 1h20 & 0h35 & 0h25 & 1h30 & \textbf{11h15} \\ \hline
         \textbf{E4} & 8h00 & 2h30 & 2h20 & 2h30 & 0h30 & 2h00 & \textbf{17h50} \\ \hline
         \textbf{E5} & 7h00 & 2h00 & 2h55 & 1h30 & 1h30 & 1h00 & \textbf{15h55} \\ \hline
         \textbf{Ave.} & \textbf{6h24} & \textbf{1h19} & \textbf{2h31} & \textbf{1h12} & \textbf{0h44} & \textbf{2h00} & \textbf{14h10} \\ \hline
    \end{tabular}
    \caption{Time spent by each expert in learning and in each task}
    \label{tab:timeAnalysis}
\end{table}

Question \textbf{Q9} asked experts to rate the approach, which received an average rate of 9. Question \textbf{Q11} asked experts to analyze each contribution. The average ratings were the following: C1=9.2, C2=10, C3=8.8, C4=9, C5=9.2 and C6=8.4. Overall, all experts provided very positive results for all questions, which might represent some bias and is a possible validity threat, as discussed later. But it also indicates that these contributions are being achieved, in particular \textbf{C2}, which received maximum rating from all experts. Some smaller values were provided by some experts in particular cases. These are analyzed in the following sections.

\begin{figure}[htb]
\begin{center}
\includegraphics[width=1\linewidth]{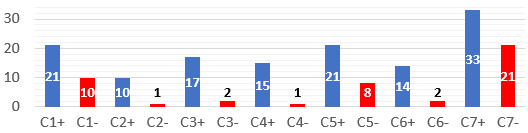}
\caption{Number of annotations in favor (\textbf{C+}) and against (\textbf{C-}) each contribution.}
\label{fig:graficoContribuicao}
\end{center}
\end{figure}

Figure \ref{fig:graficoContribuicao} shows the total amount of annotations for each contribution, after an analysis of the responses. As it can be seen, in general, all contributions have a positive result, i.e. with more annotations in favor than against them.

Although the quantitative results are overall positive, there were limitations and negative observations made by experts. In the following section each contribution is individually analyzed, based on the expert responses.

\subsubsection{Qualitative analysis}


A total of 21 annotations favor contribution \textbf{C1 (High abstraction level modeling in a single environment)}. 33\% of the annotations regarding \textbf{C1} appear in question \textbf{Q2}. Four experts (\textbf{E1}, \textbf{E2}, \textbf{E3} and \textbf{E4}) were able to identify the higher abstraction levels and hiding of technical details behind the platform models. Among these 21 annotations, the following excerpts from the transcript exemplify this perception:

\textbf{E2}: \say{The complexity is low because most details are concentrated in the platform models (...) It is possible to keep the logic in a high level, using the GPL.}. \textbf{E3}: \say{The approach increases the abstraction level and considerably reduces the complexity}.

There were also 10 noticed limitations for contribution \textbf{C1}. Four experts (\textbf{E1}, \textbf{E2}, \textbf{E3} and \textbf{E5}) reported the lack of a way to test the algorithms written in the GPL. This is not a problem with the approach, as it is possible to create a platform for generating test code, but the experts failed to see this possibility. Expert \textbf{E2} also emphasized, in question \textbf{Q2}, that the GPL could be improved with syntax validation and code completion. There are some basic functions in this regard provided by Xtext, but improvements are necessary to make the current implementation more usable.


Another limitation pointed out by the experts (\textbf{E2} in \textbf{Q6} and \textbf{E5} in \textbf{Q10}) is a lack of confidence in the generated code. They did not point out any error during the experiment, but they raised the question, which indicates that they might not trust code generation. Expert \textbf{E2} also pointed out possible performance problems in the generated code. Although most of these are probably solvable through GPL programming and platform optimizations, the concern reinforces this lack of trust. This has a negative impact on contribution C1, as the developer will have to inspect and test the lower abstraction level models and generated code to be more confident.

There are two sources of possible errors in code generation: in the code generators and in the platform models. The code generators might have some problems, but they should disappear as the approach is used more often and bugs are discovered. The platform models may also introduce errors, and they must be tested to increase confidence. As pointed out by \textbf{E5}, the approach does not provide a good support for native programming, and the developer will have to resort to the native IDE. This also goes against C1.

Expert \textbf{E1}, in \textbf{Q10}, reported difficulty in creating some of the classes for the ``controller'' layer. Having to customize these classes for each platform, through the concept of global details, required a new way of thinking. According to the expert, this was difficult to grasp at first, but once the concept is fully understood it may lead to long term benefits.

Expert \textbf{E4} reported in \textbf{Q10} the lack of support for the ``View'' layer. This was expected, as this layer not included in the platforms provided for the study.

According to expert \textbf{E3} in \textbf{Q10}, the higher abstraction level would benefit from a visual representation of the classes. Since the GPL is a textual language, developers might find it difficult to understand the relationship between the classes. This is also a limitation related to C1.


Now we analyze contribution \textbf{C2 (Reuse of similar concepts between platforms through modeling)}. There were 10 annotations in favor of this contribution, with around 30\% appearing in \textbf{Q4}. These are indications that the approach can facilitate the reuse of similar concepts between platforms. Some examples extracted from the transcript illustrate the experts' opinions: \textbf{E1}: \say{(The approach) can use the same algorithm to generate code for different platforms.} \textbf{E2}: \say{If the GPS example worked for Android and iOS then it works with any other resource.}

In question \textbf{Q7} the experts pointed out advantages that come from reusing the source code in different platforms during system maintenance, with changes being performed at high level in the GPL (\textbf{E1} and \textbf{E4}). For example, expert \textbf{E4} stated that \say{All you have to do is save the changes and the native code for every platform is updated.} Experts also mentioned the possibility to reuse generic algorithms for different platforms in \textbf{Q5} (\textbf{E2} and \textbf{E3}).

The reuse of similar concepts between platforms, however, was perceived as a medium to long term benefit, as it is necessary to develop the platform models first. For example, \textbf{E3}, in \textbf{Q2}, stated: \say{I believe the benefits will only be reached in the long term, when all the platform models and global functions are implemented}. According to \textbf{E3} and \textbf{E5}, having to manually create and test, in their native IDEs, all the global functions for a cross-platform system requires a lot of effort. This initial effort can be reduced as new systems reuse existing models. Importing older models was cited as a possible way to overcome this limitation, as mentioned by some experts (\textbf{E5} in \textbf{Q7}, \textbf{E1}, \textbf{E3} and \textbf{E4} in \textbf{Q6} and \textbf{E5} in \textbf{Q4}. Expert \textbf{E1} even suggested that a collaborative repository could be made available to facilitate reuse. 


Contribution \textbf{C3 (Reduced system configuration costs)} was unanimously reported in \textbf{Q5} as a positive contribution. Experts pointed out some advantages in using a model specifically to configure code generation:

\textbf{E1}: \say{It is possible to easily see where each class is directed (which platform) and manage functionality distribution}. \textbf{E3}: \say{I found it very simple, very interesting, since we can leave everything in a single block and more easily manage the multiplatform system}. 

However, some limitations were pointed out. As stated by \textbf{E3} in \textbf{Q2} and \textbf{Q8}, and \textbf{E5} in \textbf{Q5}, the benefits of the deployment model depend on the existence of implementations for the global functions: \textbf{E3}: \say{The approach separates the development among the various abstraction levels involved in the process. However, I believe the gains will be achieved in the long term, when the platform models and global functions are implemented}.

Another limitation associated with \textbf{C3} was highlighted by \textbf{E3} in \textbf{Q3}, regarding large systems and library dependencies. In these systems, libraries may become deprecated or suffer changes that require the platforms to be constantly updated. In the studies no such problems appeared, but the fact that an expert demonstrated concern in such scenarios is a potential limitation that must be addressed.


Annotations regarding contribution \textbf{C4 (Inclusion of new platforms with a small impact)} appeared predominantly in questions \textbf{Q3} and \textbf{Q5}. According to the responses to these questions, the approach supports the inclusion of any new function, and any platform that can be developed in one of the supported languages can be included. For example, \textbf{E2} says: \say{I think it is possible to add any technical function in the platform models}.

Regarding the impact of such inclusion, experts \textbf{E1}, \textbf{E4} and \textbf{E5} in \textbf{Q5} report that it is low, as long as there are some models and implementations from other platforms to be reused. If a new platform has to be completely developed from scratch, there will be considerable effort, as reported by expert \textbf{E1} in \textbf{Q11}. Knowing how to explore reuse is a key factor, according to this expert.


In order to analyze contribution \textbf{C5 (Extension of the approach through platform models)}, we asked experts to extend one of the platforms in Task \textbf{T3}, to give them the basis to form an opinion. All experts stated that the approach allows the evolution of the platform models by themselves, without having to wait for new releases from a third-party tool vendor. Another point of interest, made by \textbf{E2} and \textbf{E3} in \textbf{Q3} is the fact that the developer can adjust the generated code. This is an advantage over existing frameworks, as discussed earlier. For example, \textbf{E2} says: \say{Flutter generates code that cannot be changed. Your approach does not have such problem, since it is possible to extend platform models and work with a readable generated code}.

This particular comment from \textbf{E2} is significant, as it is one of the main benefits expected for the approach. In his comment, the expert cited some problems regarding the differences in configuration and permissions for using the camera and GPS in the iOS and Android platforms. According to \textbf{E2}, Flutter does not allow fine adjustments in each platform, treating both platforms in the same way. He also cites specific requirements for publishing applications in the app stores, which also may require adjustments on the generated code for each platform. Since it is not possible to modify the generated code with Flutter, this may cause hard to solve problems during the execution or publication of the application. This approach, on the other hand, allows these changes to be made by the software engineer on the platform models.

The negative comments on this contribution refer to the lack of trust on the generated code. According to \textbf{E2} in \textbf{Q10}, there must be some way to guarantee that the approach generates code that is correct and with good performance. As discussed in \textbf{C1}, the software engineer has control over most of the generated code. Although this brings flexibility, it has a downside, since she is also responsible for testing it and make sure the code is working as expected.

Another limitation mentioned by \textbf{E1}, \textbf{E2} and \textbf{E3} in \textbf{Q10}, and \textbf{E3} and \textbf{E5} in \textbf{Q3} is the absence of a test mechanism for new functions in the platform models. This is a negative factor for \textbf{C5}, and is similar to the limitation mentioned regarding \textbf{C1}. But here the problem is different, since it refers to testing the platforms, and not the system. Ideally, this should be made in the native language, using a native IDE, which goes against the purpose of using a single environment to create and evolve the software. On the other hand, simpler adjustments and extensions can be made in the models, as reported by \textbf{E4} in \textbf{Q8}.


The annotations regarding \textbf{C6 (Code reuse across domains)} appear in \textbf{Q3} (35\%) and \textbf{Q4} (65\%). In \textbf{Q3}, some experts reported that the \textit{Apache Velocity} templates help to make the platform models generic enough to be used in other domains. In \textbf{Q4} all experts reported that the generated code can be fully customized through global functions, and this allows the platform models to be used with any entity and in any domain. The following excerpts from the transcript illustrate their opinions: \textbf{E2}: \say{I believe it is possible to reuse the code and it is very practical, since it is generic. You can change details very easily, just informing which entity is to be persisted}. \textbf{E5}: \say{Global functions make a lot of sense in a multiplatform scenario, but it still makes sense for a single, specific platform}.

Expert \textbf{C4}, in \textbf{Q10}, also pointed out the limitation that the reuse was restricted to the ``Model'' and ``Controller'' layers of MVC. This was expected, as in the studies the ``View'' layer was left out of scope. Although in theory this could be implemented as a new platform and a set of global functions/details, we believe there are many details that need to be addressed, which is why this is left to future work.


The last contribution \textbf{C7 (Development and maintenance cost reduction)} was the most cited by experts in their responses, mostly because this is a generic statement that may be influenced by all other contributions. It was even cited as one of the main benefits of the approach by experts \textbf{E1} and \textbf{E3} in \textbf{Q8}.

The annotations regarding \textbf{C7} are also generic, and experts associate cost reduction with gains in productivity, management and time. The main reasons for such gains come from: the GPL (\textbf{E4} and \textbf{E5} in \textbf{Q2}); the reuse of platform models (all experts in \textbf{Q4}, \textbf{E5} in \textbf{Q2} and \textbf{Q5}, \textbf{E4} in \textbf{Q5} and \textbf{E2} in \textbf{Q3}; the overall architecture (all experts in \textbf{Q6}); facilitated maintenance (all experts in \textbf{Q7}).

The following examples illustrate the perceived benefits in cost reduction: \textbf{E5}: \say{If the global functions are implemented, productivity is greatly increased}. \textbf{E3} in \textbf{Q8}: \say{It also improves the management of the developers' roles in the process}. \textbf{E4}: \say{Time and management gains seems good}.

Overall, the opinions strongly support this contribution. Some limitations were cited, however. These appear in all questions, but most are in \textbf{Q10}. Most were already discussed in the previous sections: the lack of support for the ``View'' layer, for example, is one of them. The extra effort needed for implementing the platform models was also cited. Experts \textbf{E1}, \textbf{E3} and \textbf{E4}, in \textbf{Q6}, explicitly state that the benefits are only possible under the condition that all platform models are implemented. Expert \textbf{E5} made comments in this regard in almost all questions. Expert \textbf{E3} also points out the need for extra testing for the platform models.

Another limitation is that the approach requires experts in different languages for the proper implementation of all global functions for all platforms (\textbf{E3} in \textbf{Q4}). However, the same expert indicates that, in the end, this pays off in terms of productivity gains (\textbf{Q4}) and better role management (\textbf{Q8}).

\subsection{Novice Evaluation}
\label{sec:ThirdEvaluation}

The third evaluation was similar to the second, but it involved novice developers, not familiar with web and mobile development. We selected four undergraduate students from the last year of their System Analysis and Development course offered at Federal Institute of São Paulo (IFSP) - campus Piracicaba - Brazil. These students had basic programming knowledge and minor experience with IDEs, but they still hadn't studied mobile and web development. Since the goal of this evaluation was to test if these students could develop software for web and mobile platforms, all the necessary platform models were provided to them prior to the study. As in the previous study, participants were asked to perform some tasks for the e-commerce domain, but here the tasks did not require understanding or modifying the platform models (global functions and global details):


\begin{description}
\item[Task 1:] create an authentication function for a mobile app.

\item[Task 2:] create a function to retrieve products stored in SQLite database to present to the end user.

\item[Task 3:] support for updating products in a shopping cart. 

\item[Task 4:] reconfigure how each part of the application is deployed in the different platforms, and use GPS location in some specific platforms.

\end{description}

All tasks required creating classes for the ``Model'' and ``Controller'' layers, with business logic being specified using the GPL. Reuse of global functions and global details was expected. As in the previous study, the ``View'' layer was already provided and did not require development.

This evaluation also had three stages, as the previous one. A repository with all necessary material (documentation, tutorials and necessary tools) was made available for the participants\footnote{The complete novices material and transcripts are available at:  \url{https://github.com/JulianoZ/CrossPlatformApproach}}. The same interview questions used in the previous study was used here, but questions 3 and 4, which focused on understanding and modifying platform models, were not considered as relevant because beginner participants were not expected to be able to answer them properly. All interviews were transcribed and analyzed by the two researchers, as in the previous study.

\subsubsection{Quantitative analysis}

All tasks were successfully completed by the novice participants. Table \ref{tab:timeAnalysis2} presents the time spent by each one. The average time was 16h48min, with an average learning time of 9h15min. These values were higher than those spent by the experts (Table \ref{tab:timeAnalysis}), even considering that experts had more tasks to do (modifying platform models). This is expected due to the lack of experience of these participants.

\begin{table}[htb]
    \scriptsize
    \centering
    \begin{tabular}{|c|c|c|c|c|c|c|c|c|} \hline
         \textbf{Nov.} & \textbf{Learn.} & \textbf{T1} & \textbf{T2} & \textbf{T3} & \textbf{T4}  & \textbf{Total} \\ \hline
         \textbf{N1} & 10h00 & 2h30 & 1h00 & 3h00 & 0h30  & \textbf{16h40} \\ \hline
         \textbf{N2} & 12h00 & 3h00 & 1h30 & 3h00 & 1h00  & \textbf{20h30} \\ \hline
         \textbf{N3} & 7h00 & 2h00 & 1h30 & 3h35 & 0h40  & \textbf{14h05} \\ \hline
         \textbf{N4} & 8h00 & 4h00 & 1h30 & 2h30 & 0h50  & \textbf{16h10} \\ \hline
         \textbf{Ave.} & \textbf{9h15} & \textbf{2h50} & \textbf{1h20} & \textbf{2h55} & \textbf{0h45} & \textbf{16h48} \\ \hline
    \end{tabular}
    \caption{Time spent in learning and in each task}
    \label{tab:timeAnalysis2}
\end{table}

The ratings provided in questions \textbf{Q9} (average=9.25) and \textbf{Q11} (C1=9.5, C2=10, C3=9.5 and C6=9.5) were positive, indicating that the approach was seen as mostly positive by novice developers, delivering the expected contributions, except for C4 and C5 which were not considered here.

\begin{figure}[htb]
\begin{center}
\includegraphics[width=1\linewidth]{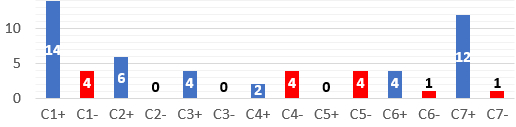}
\caption{Number of annotations in favor (\textbf{C+}) and against (\textbf{C-}) each contribution.}
\label{fig:graficoContribuicaoAmadores}
\end{center}
\end{figure}

Figure \ref{fig:graficoContribuicaoAmadores} summarizes the amount of annotations for each contribution made in the interview transcripts. It is possible to see here indications that the approach delivers most contributions, with positive annotations outnumbering the negative annotations. The exceptions were C4 and C5, with negative observations being more frequent. As detailed later, these annotations reflect the fact that participants reported not being able to understand, modify or create platform models (they were only able to reuse them). This was expected, as these participants were students without knowledge in mobile or web development, and platform models are essentially a place to put such technical details.

Another interesting observation is that, overall, the novice developers were less critic than the experts, with less negative comments. We believe this is explained by the fact that, in general, less experienced developers are also less capable of performing a deeper analysis of a new technology.

\subsubsection{Qualitative analysis}


Contribution \textbf{C1} (High abstraction level modeling in a single environment) was the most evident, as shown by the 14 positive annotations. All participants perceived an increase in the abstraction level and the isolation from technical details. Some exemplary comments follow: \textbf{N1}: ``The GPL is not complex, as it does not have technical details''. \textbf{N3}: ``It is possible to program without knowing the technical details of each platform and thus focus on application logic''.

However, there were 4 negative annotations. N3 and N4 pointed out the lack of support for the graphical user interface, and N3 pointed out the lack of resources for testing. N1 also mentioned the lack of programming resources, such as conversion and code completion. And N3 mentioned that the certainty of the code working is only possible in the native IDEs, demanding additional work. 


The positive evidence regarding \textbf{C2} (Reuse of similar concepts between platforms through modeling) corresponds to 6 positive annotations in the transcripts, with examples being: \textbf{N1}: ``I realized it can automate the creation of code for different platforms''. \textbf{N3}: ``I liked the adaptations made by the approach regarding the specifics (technical details) for different entities and platforms''.

There were no negative observations regarding \textbf{C2}, probably because the participants did not have to deal with technical details in the platform models, only reuse them, and thus everything worked. In comparison with the previous study, this was perceived differently by the experts, who observed that reuse between platforms required extra effort.


Four positive annotations regarding \textbf{C3} (Reduced system configuration costs) were made. Some examples are: \textbf{N3}: ``... it allows us to really manage which functions will be generated for each platform''. \textbf{N4}: ``You can easily choose where each system function goes''. No negative comments were made regarding \textbf{C3}.


Contribution \textbf{C4} (Inclusion of new platforms with a small impact) had more negative (4) than positive (2) comments, with the most illustrative examples being: \textbf{N1}: ``...I don't really know how to use them in other situations, or how to create new global functions''. \textbf{N3}: ``I did not understand how to create a global function''.

The participants showed concern and doubts when dealing with global functions. Although they could successfully reuse them, they wondered if they were capable of understanding or creating new ones. An interesting fact is that this was not expected from them, but they reported such concerns anyway. But there were two positive comments regarding the possibility of including new platforms: \textbf{N2}: ``...it does not limit the user to the platforms supported by the tool manufacturers''. \textbf{N3}: ``It is a nice feature, that enables the user to use and include more platforms''.


Evidence regarding \textbf{C5} (Extension of the approach through platform models) is very similar to \textbf{C4}. The lack of expertise regarding technical details hindered the participants' confidence in modifying existing platform models: \textbf{N1}: ``...I understand how to use them... ...but I would not know how to create new functions''. \textbf{N2}: ``I think it is easy to use a global function... ... but I wouldn't know how to create new functions''.


Contribution \textbf{C6} (Code reuse across domains) was seen mostly positive. For example: \textbf{N2}: ``Global functions allow the creation of code for any entity by easy configuration with generic parameters.'' \textbf{N3}: ``It is possible to access global functions in any place of the system and reuse the code''.

The single negative comment reinforces the lack of confidence in reusing these resources, which is something also mentioned by experts: \textbf{N1}: ``I don't know if this may create a problem in the reuse of global functions''.


\textbf{C7} (Development and maintenance cost reduction) was the second most identified contribution, with 12 positive annotations, being unanimously cited as a benefit by the participants. Some examples are: \textbf{N2}: ``What I find most interesting is the reduction of development cost''.\textbf{N4}: ``The great benefit is the agility in development, which end up reducing the costs''. The only negative comment was made by \textbf{N3}: ``But it is necessary to use the native environments for execution and tests, which demands specific knowledge and may reduce productivity''.

\subsection{Discussion}
\label{sec:discussion}

The first study brought evidence favoring contributions \textbf{C1}, \textbf{C2}, \textbf{C3}, \textbf{C4} and \textbf{C7}, but this was just a proof-of-concept. The second and third studies brought evidence in favor and against all seven contributions of this approach. The amount of evidence in favor is larger, which indicates that the approach is successful in achieving its goals. We also observed a difference between expert and novice developers. While experts were able to see the benefits in terms of the extension/inclusion of platforms more clearly, novice developers failed to see such benefits, mostly because this requires technical knowledge regarding the platforms.

Another interesting observation was made by a participant. This was not foreseen as an important contribution, but because the third study was conducted with students, it appeared as a side benefit of the approach, as exemplified by the following comment:

\textbf{N4}: ``At the same time you learn a new GPL language, you end up learning some Java, Swift, C\#. It contributes to the learning of other languages as well.''

This is an unexpected benefit, particularly useful for learning. Students can use the approach to see examples of how a piece of abstract code translates to different platforms.

The studies also uncovered eight important limitations. These are summarized as follows, ranked in a decreasing order of severity according to our analysis:

\begin{description}
\item[L1.] Need to resort to the native language and IDE;

\item[L2.] Benefits are only expected in the long term, as global functions need to be fully implemented first;

\item[L3.] Lack of support for testing, including tests for validating the system, the platforms and the generated code;

\item[L4.] Lack of support for the ``View'' layer in MVC;

\item[L5.] Need to understand a new programming pattern, which may have a steep learning curve;

\item[L6.] Lack of trust in the generated code correctness;

\item[L7.] Lack of trust in the generated code performance; and

\item[L8.] Lack of support for a visual modeling language.
\end{description}

The first limitation is a more fundamental one. Initially, the approach was supposed to support a single environment for creating the software, but in the end it became clear that the native platforms' IDEs have to be used when creating platform implementations. As the time passes and the platforms are implemented and tested, there is less need to resort to the native IDEs. This is also the essence of limitation \textbf{L2}.

\textbf{L3} and \textbf{L4} are also important, but we think they can be solved by providing new platforms specifically for these tasks. Regarding \textbf{L3}, we conducted a fourth evaluation specifically for this reason, as described later. Regarding \textbf{L4}, global functions and details may be used to represent common interface concepts, such as event handling and state management. It might even be possible to create platforms for domain-specific interfaces, based on the fact that many systems from the same domain share similar interfaces. These are the subject of future work.

\textbf{L5} is also mentioned as an important limitation. The concept of global functions and details requires learning a new programming technique.

\textbf{L6} and \textbf{L7} are the other side of the flexibility provided by the approach. The software engineer has more control over the generated code, but is also responsible for making sure it is correct and adequate.

Finally, \textbf{L8} was cited by one of the experts as a possibility for improvement. This limitation was not perceived in practice, however, since the textual language was acknowledged as intuitive and easy to understand.

\subsubsection{Threats to validity}

There are some aspects that must be discussed regarding possible validity threats in the studies.

The first study was carried out by the researchers, what might have introduced bias. To reduce this bias, no opinions or data related to the learning curve were taken into consideration in this study. Only the feasibility of the approach, its models, and the ability to include a previously unsupported platform were tested in the proof-of-concept.

For the second and third studies, there are also some identified threats. Finding novice developers was easier, as the researchers also work in the academia and are constantly teaching software development subjects to inexperienced developers. But finding experts was a very difficult task, as it required many hours of volunteer work. The researchers resorted to their contacts in academia and industry and it was possible to find five experts that fulfill the requirements (minimum 10 years of experience with mobile and web development and some experience with multiplatform development). However, this sampling was not ideal as the participants knew the researchers and might have provided biased answers. Nevertheless, the amount of negative observations made by the experts was considerable and indicates that the bias might have not been very strong. It also indicates that their level of expertise was adequate enough to allow them to identify real problems, going beyond a superficial analysis.

It was also very difficult to define tasks for the participants. The entire study took an average of 14 hours of work from each expert and 17 hours from each novice developer. These were not performed uninterruptedly, as the participants worked a small amount of time each day. It took several days to complete all tasks, therefore they were in contact with the approach for a long time. Ideally, the tasks should be longer, so that the participants could have a deeper understanding of all its details. However, this was not possible because their availability for the study was partial.

Still regarding the tasks, we attempted to cover different scenarios to exercise the different contributions of the approach. For example, task T3 asked experts to modify an existing platform to support a new database, and T5 asked experts to include a new platform (Android Wear). In the end, it was possible to collect evidence regarding all contributions. However, no task required an entire system to be developed from scratch. This could probably lead to more identified benefits and limitations but, again, was not possible due to the participants' partial availability.

Another threat is related to a possible external factor. All participants identified productivity gains, but we think that part of these gains might have originated from the \textit{Apache Velocity} templates, and not the approach. Although the templates alone do not promote cross-platform reuse and development, they can greatly reduce programming effort, specially in the CRUD domain. There were some tasks that did not involve the use of such templates, they might have influenced the participants' responses, who attributed 100\% of the productivity gains to the approach, while in reality a fraction of that probably resulted from \textit{Velocity}.

One of the limitations (\textbf{L5}) relates to the learning curve regarding the approach, in particular the use of global functions and details. Both experts and novice developers managed to complete the tasks, but they had examples to follow, therefore the learning aspect was not completely addressed.

Finally, the second and third studies had strong subjectivity, in two aspects. Personal opinions are subjective. To reduce this subjectivity, we selected five experts with experience, to increase the confidence that their opinions are not isolated or based on guesswork. The second subjective aspect was the analysis of the participants' opinions, carried out by the researchers. To reduce this threat, the two authors of this paper conducted the analysis separately, and then a consensus was reached after discussion.

\subsection{Fourth evaluation: unit testing}
\label{sec:fourthEvaluation}

This fourth study was conducted by the researchers as a response to observations made by experts and novice developers regarding testing in the platforms (Limitation \textbf{L3}). After discussions, we thought this could be done by creating platform models for unit testing in different platforms. In this solution, global functions are created to represent assert functions, present in most unit testing frameworks. Each platform model is responsible for defining how the global assert functions translate to its own platform-specific assert functions, including additional code, such as annotations, if necessary, as global details. Test cases can reuse the global functions and call the assert functions normally.

Listing \ref{lst:testClass} illustrates the solution. By reusing a global assert function (line 2) and including specific details required by unit test frameworks (line 3), the developer can simply call the assert function (line 21) normally. In this example, an error was deliberately introduced in the ``Debit'' method (line 16) to cause this test case to fail.

\begin{lstlisting}[caption={Cross-platform test case using the approach.},label={lst:testClass}]
class BankAccountTest{	
	usesGlobal  globalAssertEquals(expected: double, actual: double, msg: string): void{}
	usesGlobalDetails UnitTestAnnotation 
	operation balanceTest(): void{
		m_customerName: string
		m_customerName := ``Juliano''
		beginningBalance: double
		debitAmount: double
		expected: double
		beginningBalance := 11.99
		debitAmount := 4.55
		expected := 7.44
		account :BankAccount
		account.setM_customerName(m_customerName)
		account.setM_balance(beginningBalance) 
		account.Debit(debitAmount)
		actual : double
		actual := account.getM_balance()	
		msg: string
		msg := ``Account not debited correctly''
		global globalAssertEquals(expected, actual, msg)
}   }	
\end{lstlisting}

\begin{figure}[htb]
\begin{center}
\includegraphics[width=1\linewidth]{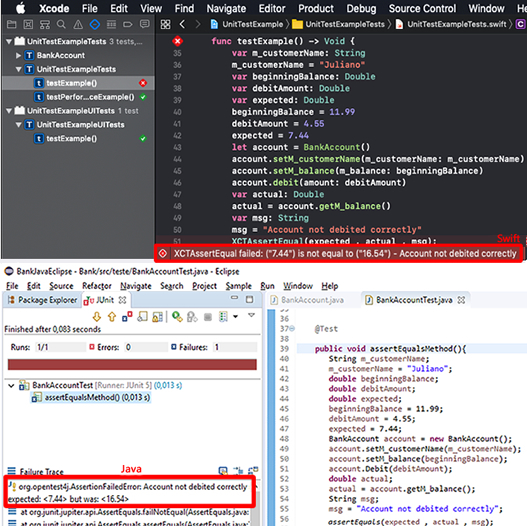}
\caption{The same test case executed in iOS and Android.}
\label{fig:UnitTesting}
\end{center}
\end{figure}

After properly configuring code generation for each desired platform, the generated code was successfully executed in the different platforms, as shows Figure \ref{fig:UnitTesting}. In the figure, the test case from Listing \ref{lst:testClass} is being executed in iOS and Android, with the same error being accused in both cases.


This evaluation demonstrates that the approach can be used to test application logic in multiple platforms, but there are other issues to be explored, such as testing native resources and integration tests. These are left for future work.

\section{Concluding remarks}
\label{sec:conclusion}

Cross-platform development solutions have been investigated for a long time, and new research is often arising in the industry and academia. There are many open areas for research, as existing solutions are still not mature enough \cite{chen2019automated}.

In this spirit, this paper presents an approach for cross-platform development, where the main idea is to keep development in a single, platform-independent modeling environment, and use generators to produce executable code. The approach takes advantage of similar concepts across platforms and facilitates the inclusion and exchange of platforms used by the system. As a consequence, developers can benefit from reuse provided by code generators and still have control to customize, extend or include new platforms.

The approach targeted seven contributions, selected after an analysis of the solutions available in industry and academia. Four studies were conducted to evaluate them. There were some threats to the validity of the studies, but overall the collected evidence supports the contributions. In particular, the studies demonstrated the ability to extend or modify existing platforms, what is not possible in most existing tools. We also created a new type of device and included it into the approach successfully. In most existing tools, these tasks would be impossible, as the modification of platforms and the inclusion of new platforms depend on the tool vendor, and not the developer. The studies also demonstrated the distribution of functionality among platforms, what is also not a common feature in most existing tools. These features give the approach a holistic view over cross-platform development.

The studies have also shown that it is not currently possible to completely abandon the platforms' native IDEs. This observation goes against the approach's primary goal of using a single environment to create the software. This problem tends to be reduced as time passes and the platforms become more stable, with benefits in the long term.

Future work is being planned to overcome the limitations identified in the studies, as described in Section \ref{sec:discussion}. In particular, we plan to experiment with the ``View'' layer, as it was not considered in the research so far and was pointed out by most participants of the studies as a missing feature.

Some practical improvement possibilities were also identified. One of them is the use of Object Oriented Programming concepts in the platform models, allowing inheritance between platform models and facilitating reuse and system management. One of the experts also suggested a model repository. While it is possible to reuse models in the approach by copying the files, more elaborated ways of reuse could be the subject of future work, so that platform models may be reused within a community. Similarly to code repositories associated with dependency management systems, this effort could help to reduce the initial effort mentioned by experts when using the approach.

Finally, while our approach gives expert developers the possibility to deal with all complexities to each platform, novice developers would be more comfortable if they could be somehow shielded from this complexity, as it is common in most commercial tools. This is also subject of future work.

\section*{Acknowledgement}

The research work reported in this paper received financial support from grants \#2015 / 24429-1 and \#2017 / 25343-9 - São Paulo Research Foundation (FAPESP) - Brazil.

\printcredits

\bibliographystyle{cas-model2-names}

\bibliography{cas-refs}


\bio{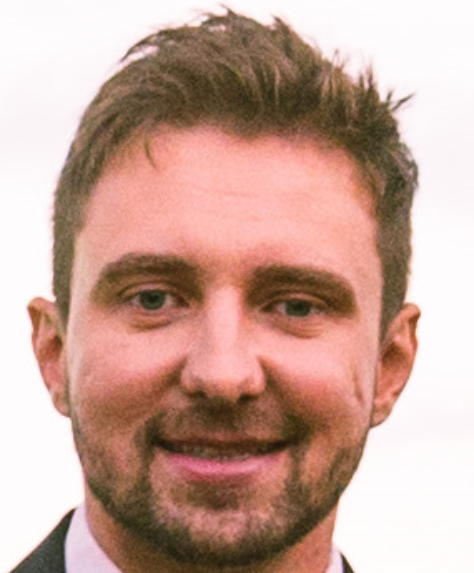}
{\textbf{Juliano Zanuzzio Blanco}} is graduated in Data Processing by Americana College of Technology (FATEC-2002), M.Sc. (2009) and PhD (2020) in Computer Science degrees at the Federal University of São Carlos, Brazil. Currently belongs to the faculty of the Federal Institute of São Paulo Campus Piracicaba (IFSP). Has experience in Computer Science, focusing on web and mobile application software development.

\endbio

\bio{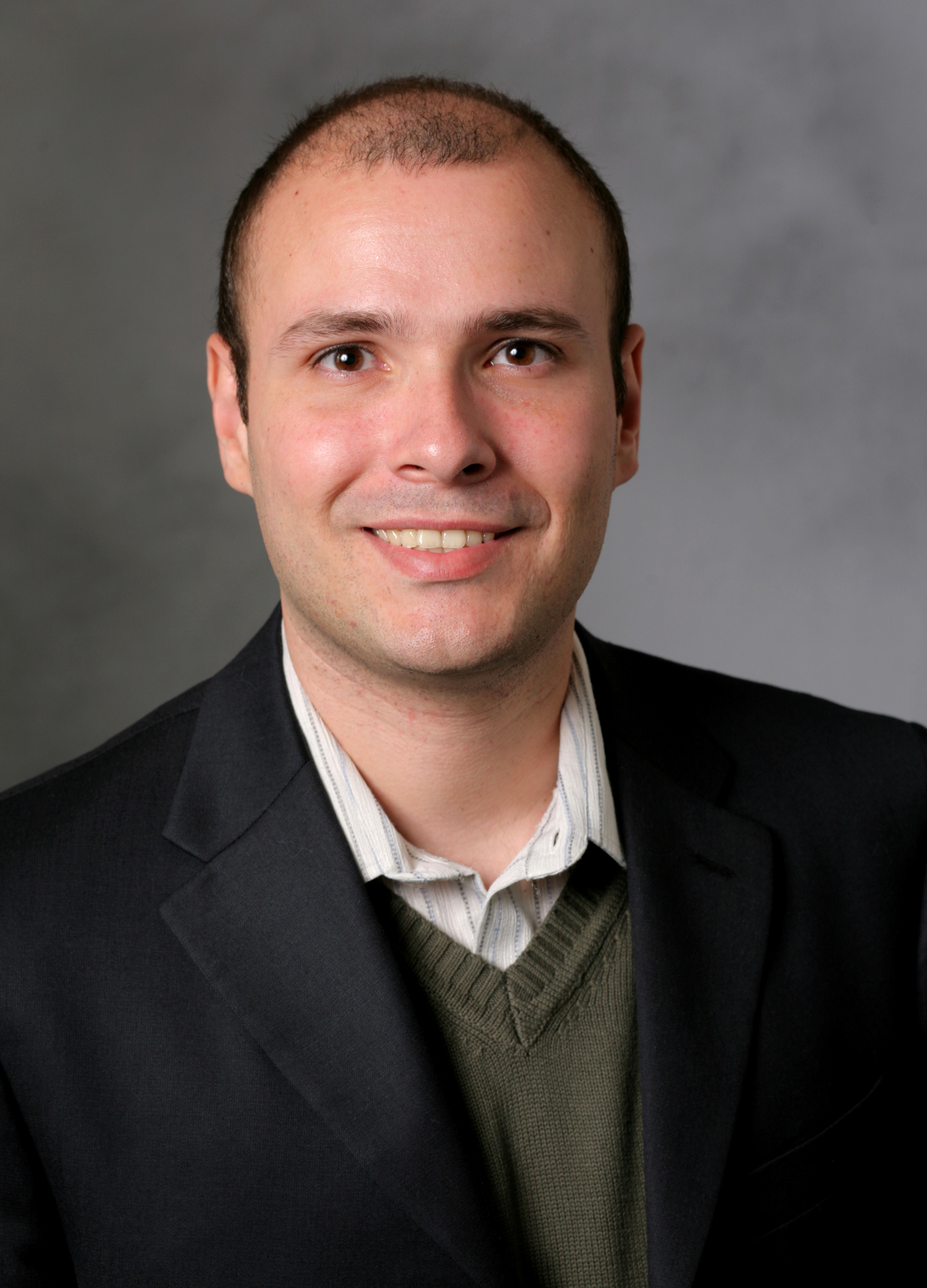}
{\textbf{Daniel Lucrédio}} got the Computer engineer (2002) and M.Sc. in Computer Science (2005) degrees at the Federal University of São Carlos, Brazil, and PhD (2009) at University of São Paulo, São Carlos, Brazil, with two doctoral internships: George Mason University (VA, USA) and Microsoft Research (WA, USA). Currently an associate professor at Federal University of São Carlos, Brazil, working mainly with Model-Driven Engineering, Cross-platform software development and Cloud computing.
\endbio

\end{document}